\documentstyle[twocolumn,aps]{revtex}

\begin{document}

\draft

\title{Thermal Conductivity of the Spin Peierls Compound CuGeO$_3$}
\author{Yoichi Ando,$^{1}$ J. Takeya,$^{1}$ D.L. Sisson,$^{2}$ 
S.G. Doettinger,$^{2}$ I. Tanaka,$^{3,4}$ R.S. Feigelson,$^{3}$ 
and A. Kapitulnik$^{2}$}
\address{$^{\rm 1}$ Central Research Institute of Electric Power
Industry, Komae, Tokyo 201, Japan}
\address{$^{\rm 2}$ E.L. Ginzton Laboratory, Stanford University, 
Stanford, CA 94305}
\address{$^{\rm 3}$ Center for Materials Research, Stanford University, 
Stanford, CA 94305}
\address{$^{\rm 4}$ Institute of Inorganic Synthesis, 
Faculty of Engineering, Yamanashi University, Kofu, Yamanashi 400, Japan}
\date{Received CuGeO-6.tex}
\maketitle

\begin{abstract}
The thermal conductivity $\kappa$ of the Spin-Peierls (SP) compound 
CuGeO$_3$ was measured in magnetic fields up to 16 T.  
Above the SP transition, 
the heat transport due to spin excitations causes a peak at $\sim$22 K,
while below the transition the spin excitations rapidly diminish and 
the heat transport is dominated by phonons;
however, the main scattering process of the phonons is 
with spin excitations, which demonstrates itself in an unusual peak in 
$\kappa$ at $\sim$5.5 K.  
This low-temperature peak is strongly suppressed with magnetic fields 
in excess of 12.5 T.
\end{abstract}



The recent discovery of the inorganic Spin-Peierls (SP) compound 
CuGeO$_3$ \cite{Hase} has revived interests in this phenomenon 
\cite{Cross,Bulaevskii}. 
Spin-Peierls transition occurs in quasi-one-dimensional magnetic systems 
when the energy gained by splitting the degeneracy of the original 
magnetic ground state exceeds the lattice deformation energy. 
The SP ground state is therefore a nonmagnetic singlet state 
with a magnetic energy gap between the singlet and the lowest excited 
triplet states. 
The appearance of the magnetic energy gap and the accompanying dimerization 
of the $S$=1/2 Cu$^{2+}$ ions have been confirmed 
by magnetic susceptibilities \cite{Hase}, inelastic and elastic neutron 
scatterings \cite{Nishi,Regnault,Pouget,Hirota}, X-ray diffractions 
\cite{Pouget}, and other experimental probes. 
The size of the spin gap has been found to be 
$\Delta\sim$23 K \cite{Nishi}. 
This gap can be suppressed with magnetic fields; 
above a critical field $H_c\simeq$12.5 T, the system undergoes 
a first-order transition to an incommensurate (I) phase \cite{Hase2}. 
The incommensurate lattice pattern has been measured by X-ray experiments 
and interpreted as soliton lattice structure \cite{Kiryukhin}. 

Although CuGeO$_3$ shows clear evidence for SP transition, 
there has been accumulating evidence that CuGeO$_3$ is not 
a prototype SP system. 
For example, inelastic-neutron-scattering \cite{Nishi} find 
the nearest-neighbor (NN) interaction $J\sim$120 K 
and a relatively strong interchain coupling ($J_b\sim 0.1J$). 
Static susceptibility, which should peak at $k_{\rm B}T\sim J$ 
for an ordinary SP system, shows a broad maximum centered at a temperature 
much smaller than $J$.
To describe the magnetic properties of this system,
a model which includes next-to-nearest neighbor (NNN)  
interaction $\alpha J$ has been proposed \cite{Dorby,Castilla}.
Since both NN and NNN interactions have the same sign, 
large $\alpha$ is a source for frustration. 
For $\alpha < \alpha_c$ ($\alpha_c$ estimated as 0.241 \cite{Castilla}) 
the ground state of the magnetic system remains gapless,
while for $\alpha > \alpha_c$ the spectrum of the spin waves becomes gapped
even above the SP transition temperature $T_{SP}$. 
Recent estimates of the frustration parameter suggests $\alpha$ 
to be as high as 0.36 \cite{Dorby}.

While the phase diagram of CuGeO$_3$ is now well established, 
no transport measurements have been reported to date on this system. 
Such measurements are important for the determination of 
the basic scattering mechanisms in this system, both above and below 
the SP transition and in the high-field I phase. 
Since the SP transition is basically a magnetoelastic transition, 
heat transport which is expected to take place by phonons and 
spin excitations is ideal to probe these quasiparticles.  

In this Letter we report thermal conductivity measurements 
of high quality CuGeO$_3$ single crystals along the chain direction 
\cite{chain} in magnetic fields up to 16 T. 
One of our main results is an observation of the spin-excitation continuum
above the SP transition causing a broad maximum 
at $\sim$22 K in the thermal conductivity $\kappa(T)$.
This spinon heat transport disappears rapidly below $T_{SP}$,
causing a kink in $\kappa (T)$,
and phonons become the dominant heat carrier. 
The most intriguing observation is that the phonon thermal conductivity 
below $T_{SP}$ shows an unusual peak at $\sim$5.5 K.
This peak is strongly suppressed with magnetic field 
above $H_c \simeq$12.5 T where the SP order disappears.
In the I phase above $H_c$ the thermal conductivity 
increases, which supports the idea of a new spectrum of magnetic 
excitations that are mobile and can carry heat through the crystal. 

The thermal conductivity results are supplemented with specific heat and 
magnetic susceptibility measurements all done on the same samples. 
The specific heat data are useful for deriving the thermal diffusivity
(and thus the information on scattering times) from $\kappa$;
in the analysis of the low-temperature specific heat data, 
we suggest a new fit that is consistent 
with the simple theoretical idea of an almost 
mean-field SP transition and try to resolve some controversy over 
the analysis of the specific heat data \cite{Liu_Comment}. 

Single crystal fibers of CuGeO$_3$ were grown using the laser-heated 
pedestal growth technique, a miniaturized float zone process \cite{Feigelson}.
Powders of pure CuGeO$_3$ are cold-pressed and sintered into a pellet, 
from which we cut several source rods of about 2.6 mm in diameter.
Previously grown single crystal fibers are used as seeds. 
Our samples were grown using a 20 mm/h pull rate (seed) and about 5 mm/h 
feed rate (source rod) in a flowing O$_2$ atmosphere. 
The as-grown crystals are $\sim$4 cm long and 1 - 2 mm in diameter.

Thermal conductivity measurements were performed using 
a \lq\lq two thermometer, one heater" method. 
Samples of size $7\times 2\times 0.2$ mm$^3$ were cut from the as-grown crystal, 
where the long dimension is along the chains ($c$-axis). 
The base of the sample was anchored to a copper block held at 
the desired temperature. 
A strain-gauge heater was used to heat the sample. 
A matched pair of microchip Cernox thermometers were carefully calibrated 
and then mounted on the sample. 
The maximum temperature difference between 
the sample and the block was 0.2 K. 
Typical temperature gradient used between the two thermometers 
was $\sim$0.5 K above 20 K and 0.2 K below 20 K. 
Radiation shield was used and was kept at the average 
temperature of the sample.
Specific heats were measured in the temperature range 1.7 - 20 K 
using the relaxation method.  
The addenda heat capacity was measured independently, 
with no sample attached, in separate runs.  
The particulars of our apparatus, which allows us an extremely
high sensitivity measurements, are described in greater detail 
elsewhere \cite{Moler}. 
Magnetic susceptibility measurements were performed using a Quantum-Design 
SQUID magnetometer in the field range of 0 - 7 T.

Figure 1 shows the thermal conductivity of a CuGeO$_3$ 
single crystal in the temperature range 1.5 - 40 K. 
We also plotted in Fig.1 the specific heat data measured on the 
same crystal in the temperature range 1.8 - 16 K. 
Note that the transition is clearly seen in the specific heat 
and its sharpness is similar to the best crystals reported to date. 
Before we turn to the discussion on the thermal conductivity, 
it is useful to analyze the specific heat data 
and see its correspondence to previously published results. 
Ignoring the transition region, 
we fit the data above the SP transition 
($T_{SP}$=14.08 K) to the form: 
$c/T = \gamma + \beta T^2$, 
while at low temperatures we use the form: 
$c/T = \delta (\Delta/k_{\rm B}T)^n e^{-\Delta/k_{\rm B}T} + \beta T^2$.
Note that we introduced the ratio $(\Delta /k_{\rm B}T)^n$ as 
a prefactor to the exponential decay of the spin contribution 
at low temperatures. 
While all previous publications \cite{Liu,Lorenz} used $n$= 1,
such a fit is erroneous and will hold only at very low temperatures, 
typically below 1\% of the transition temperature. 
At higher temperatures an exact expression for $c/T$ of the spin system 
is needed for a meaningful fit.  
In fact, we believe that the discrepancies in the literature 
\cite{Liu_Comment} between 
the value of $\beta$ above and below $T_{SP}$ stems 
from the use of incorrect fitting formula for the low temperature portion.
Since the actual transition is 3-dimensional in nature, 
the most reasonable model to use is that of BCS which gives $n$=5/2 
\cite{Abrikosov}. 
While the fit above the transition is straightforward, 
at low temperatures using the full formula with the gap function is 
problematic, for the fitting program will encounter a singular matrix.  
Since the effect of the SP transition on the Debye temperature 
is expected to be negligible,  
a physical approach is to use the value obtained for $\beta$ 
above $T_{SP}$ for the low temperature fit.  

Following the above discussion, we first fit the temperature range 
above the transition (14.5 - 20 K), which gives 
$\gamma$ = 105.5 mJ/mole$\cdot$K$^2$ and $\beta$ = 0.2164 mJ/mole$\cdot$K$^4$. 
We then fit the low temperature portion (1.8 - 5 K) using the same $\beta$ 
to obtain the gap. 
Using $n$=5/2 results in 
$\delta$ = 377.86 mJ/mole$\cdot$K$^2$ and $\Delta$ = 25.7 K.  
This gives the ratio $2\Delta /k_{\rm B}T_c$ of 3.65, 
indicating a slightly stronger coupling than the 
BCS weak coupling limit of 3.52.  
It is useful to check the consistency of the parameter $\delta$
by converting it to the units of number of states 
per Cu atom, $m$. 
This is easily done by multiplying $\delta$ by $J$ 
($J$ as extracted from the literature lies in the range 80 to 150 K. 
Here we choose an averaged value of $J\simeq$115 K) and by the volume 
of a unit cell ($a\times b\times c$=1.2$\times 10^{-22}$ cm$^3$).  
For the BCS formula with $n$=5/2 we get 
$m$=2.07  (note the $\sqrt{2\pi}$ factor
in the prefactor of BCS formula \cite{Abrikosov}).  
This analysis suggests that the density of magnetic excitations is 
exactly 2 states per Cu-atom at low temperatures. 
Thus, the use of the exact expression of $c/T$ for the fit at $T < T_{SP}$ 
leads to a totally consistent analysis with a reasonable value of $\Delta$.

We turn now to the thermal conductivity $\kappa$ in zero field 
shown in Fig.1. 
As the temperature is lowered from 40 K, $\kappa(T)$ increases 
and peaks at $\sim$22 K below which it starts to drop rapidly. 
At the SP transition, $\kappa(T)$ shows a kink towards a faster drop. 
However, instead of continuously dropping (presumably exponentially), 
$\kappa(T)$ shows a minimum and
a subsequent increase to a new maximum at $\sim$5.5 K 
followed by a sharp decrease as the temperature is lowered further. 
As a first approach to understand the thermal conductivity 
we use a simple kinetic approximation. 
Assuming two types of heat carriers, phonons and spin excitations, 
the thermal conductivity can be written as:
$\kappa = \kappa_s + \kappa_p = c_s D_s + c_p D_p$.
Here the subscript $s$ denotes the magnetic (spinon) component 
and $p$ the phonon contribution. 
$c$ is the specific heat and $D$ the diffusivity 
of the relevant excitations. 
While the magnetic contribution to the specific heat 
could be fitted with a $\gamma T$ term near $T_{SP}$, 
the full description of the magnetic specific heat is 
more complicated and in fact is a matter of 
current controversy \cite{Castilla,Kuroe}.  
For the pure one-dimensional system with only NN interaction, 
the specific heat will rise from low temperatures to a broad maximum at 
$\sim J/2$. 
This maximum shifts towards lower temperatures 
as the NNN interaction increases from zero. 
In fact, current models that fit the magnetic susceptibility and 
Raman scattering data suggest $\alpha$ to be in the range 
0.24 - 0.36 \cite{Dorby,Castilla,Kuroe,Muthukumar}. 
For example, Castilla {\it et al.} \cite{Castilla} 
argued that for CuGeO$_3$ the competing exchange between 
NN and NNN interactions is moderately large 
but smaller than $\alpha_c$. 
Other authors report \cite{Dorby,Kuroe} that 
a larger value of $\alpha$ ($\sim$0.35) gives good fits to the 
experimental data.

Since the magnetic contribution to the thermal conductivity 
will be proportional to the magnetic specific heat, 
it is reasonable to assume that the peak at $\sim$22 K is 
just a reflection of this magnetic density of states. 
The fact that this peak appears at a lower temperature than 
the susceptibility (typically $\sim$55 K \cite{Hase}, 
we also measured the susceptibility on the same sample 
confirming the peak at the same temperature) is a consequence 
of the temperature dependence of the diffusivity 
$D_s \sim T^{-q}$ with $q>1$. 
Assuming that $\kappa_s$ is negligible by $\sim$100 K 
and approximating $\kappa_p$ to be constant except at very low temperatures, 
we can subtract in the temperature range 15 to 50 K 
a constant phonon background of $\kappa_p \simeq$ 0.4 W/cmK. 
The remaining contribution approximates $\kappa_s$, 
which has a peak at $\sim$22 K.
Shifting this peak to $\sim$50 K in $c_s$ requires a power $q$ of 4 or 5 
in the formula of the diffusivity. 
In fact, a simple scaling approach to estimate the spinon diffusivity 
above $T_{SP}$ yields 
$D_s \sim (J/\hbar)a^2(k_{\rm B}T/J)^{-q}$,
where $a$ is the distance between adjacent spins. 
Dividing the estimated $\kappa_s$ 
by this expression for the diffusivity yields 
an approximate contribution of the magnetic excitations 
to the specific heat. 
Using $q$=4, this magnetic specific heat shows 
a broad maximum at $\sim$50 K.  
Near $T_{SP}$ the obtained curve is linear in $T$ 
with a coefficient $\gamma^{\ast} \simeq$ 31 mJ/mole$\cdot$K$^2$
(with $J\simeq$115 K), 
within a factor of 3 of the $\gamma$ estimated 
from the specific heat data near $T_{SP}$. 
Since we omitted constants and numbers of order unity, 
it is indeed very encouraging that the two estimates are 
only a factor of 3 different from each other. 
Moreover, we subtracted a constant $\kappa_p$ to obtain $\kappa_s$, 
a very simplified approximation. 
Thus, being able to obtain a reasonable value 
for the linear coefficient of the spinon specific heat 
give us confidence that indeed the maximum of the 
thermal conductivity at $\sim$22 K is of the same origin 
as the maximum in the susceptibilty. 
In addition, the fact that this maximum is shifted 
so much to lower temperatures suggests that 
the mean free path of the magnetic excitations increases very rapidly 
as the temperature is lowered.

The SP transition is manifested in $\kappa$ 
by a sharp downward change in slope. 
This drop is clearly related to the emergence of a spin gap
and thus has the same origin as the one observed in the 
susceptibility \cite{Hase} or the strong decrease of $1/T_1$ below 
$T_{SP}$ as seen in NMR measurements \cite{NMR}. 
It is however difficult to determine the exact functional form 
of the decrease because of the subsequent increase in $\kappa$ 
and the unusual peak at $\sim$5.5 K.  
Nevertheless, if we confine ourselves to $T<5$ K, 
the data can be analyzed;
assuming that $\kappa_s$ diminishes below $T_{SP}$, 
we can assume $\kappa$ is only due to phonons
for $T \ll$ $T_{SP}$. 
Using our heat capacity result, 
we can divide $\kappa$ by the lattice specific heat, 
$c_p = \beta T^3$.
The resulting diffusivity in units of cm$^2$/sec is displayed in 
the inset to Fig.1, plotted against $T^{-1}$.
Remember that we expect the plotted $D$ to represent $D_p$ 
only below $\sim$5 K ($T^{-1}>$0.2 K$^{-1}$).  
In this temperature range it is clear that 
$D \sim T^{-1}$, a temperature dependence commonly attributed to 
a phonon scattering rate due to planar defects 
such as the strain field of dislocations \cite{Berman}.  
In fact, transmission electron microscope investigation of our crystal
found planar defects whose average distance is about 10 ${\rm \mu}$m
\cite{Marshall}.

Based on the low-temperatures result discussed above, 
we can assume a model in which $\kappa$ is dominated by phonons 
below $\sim$12 K and 
exhibiting a competition between two scattering mechanisms 
that cause the peak at $\sim$5.5 K.  
If we assume phonons are scattered both by defects and by spin excitations,
we can write
$D_p = [AT + B(\Delta /T)^r e^{-\Delta /T}]^{-1}$,
where $A$ and $B$ are constants and $r$ is a power smaller than 
$n$ (as introduced above). 
Here, $A$  can be determined from the low-temperature behavior 
(inset of Fig.1).
Dividing $\kappa$ by $c_p = \beta T^3$ again and assuming $1<r<2$  
($r$ has to be in that range to accommodate the spin-excitations 
density of states as found earlier and to give a maximum to the formula), 
we estimate $B \sim 10^{-3}$ sec/cm$^2$. 
It is easy to calculate now the phonon mean free path due to 
phonon-spinon interaction, $\ell_{p-s}$. 
Using the estimates for $B$ and $r$ and 
a sound velocity of $\sim 5\times 10^5$ cm/sec, 
we find $\ell_{p-s}(T=T_{SP}) \approx$10 ${\rm \mu}$m;
this value is compatible with the phonon mean free path for 
planar-defect scattering, $\ell_{p-d}$,
estimated from the inset of Fig.1 
($\ell_{p-d}$$\approx$20 ${\rm \mu}$m at 5 K),
which agrees with the actual planar-defect distance of
$\sim$10 ${\rm \mu}$m.

Figure 2 shows $\kappa(T)$ in the field range 0 - 16 T. 
One striking feature here is the strong suppression of the 5.5-K peak 
with magnetic field;
$\kappa$ changes by more than a factor of 3 at 5.5 K, a big
magneto-thermal effect.
This strong effect of the field on the peak is another confirmation that 
its origin is related to the magnetic excitations in the SP phase. 
Note that the position of the peak does not shift much 
as a function of temperature for magnetic fields below 12 T, 
similar to the gap that does not change much. 
The peak becomes a shoulder at 14 T and a new low temperatures upturn 
below 4 K emerges at higher fields.  
At the same time, the higher temperatures data is almost unchanged 
with the broad maximum at $\sim$22 K appearing for all fields.  
The transition itself is clearly shifted towards lower temperatures 
exhibiting similar change in slope for all fields below $\sim$12 T. 
In the inset to Fig.2 we plot $\kappa$ as a function of magnetic field 
at 4.2 K to demonstrate the behavior of $\kappa$ near the peak. 
Besides the sharp drop in $\kappa$ at $\sim$12.5 T, 
a hysteresis is found upon cycling the magnetic field. 
This effect is a clear manifestation of the first-order transition 
to the I phase at high magnetic field. 

Lorenz {\it et al.} recently reported specific heat measurements of 
CuGeO$_3$ in 16 T field \cite{Lorenz}. 
The low-temperature ($T<6$ K) behavior of their 16 T specific heat data 
is consistent with 
$c = (\beta _{ph} + \beta_{mag})T^3$
with $\beta_{mag}\sim$1.4 mJ/mole$\cdot$K$^4$. 
Since in this high field and low temperature both phonon and magnetic
contributions give a $T^3$ dependence to $c$, 
one can divide $\kappa$ by $T^3$ 
to search for different temperature dependences of the scattering times. 
Again, examining $\kappa(T)$ in 14 and 16 T,
we find that the diffusivity is proportional to $1/T$ below $T\sim$4 K.  
Since $\kappa$ at low temperatures is greatly suppressed and the
diffusivity has still the $1/T$ dependence, we may conclude that
phonons are strongly scattered by solitons in the I phase.
Also, the smallness of $\kappa$ in the I phase compared to the SP
phase suggest that solitons do not carry much heat.
Note that the solitons are simultaneously magnetic and structural 
excitations: they carry spin 1/2 and are domain walls in 
the dimerized lattice. 

Finally, we briefly discuss the increase in $\kappa$ in the I phase
from 14 to 16 T.
The $T^3$ dependence of the magnetic specific heat in the I phase
was explained by Bhattacharjee {\it et al.} \cite{Bhattacharjee} 
as the specific heat of the gapless phason modes. 
The increase of this phason with increasing field
can be the cause of the increase in $\kappa$, 
although the magnetic field dependence of the phason has not been calculated. 
Another possibility is that the solitons themselves carry heat and the 
soliton contribution is proportional to the soliton population. 
However, since there is no theory for either soliton transport or 
phonon-soliton scattering, 
we cannot check the consistency of the soliton scenario.

To summarize, our thermal conductivity measurement of CuGeO$_3$ 
revealed four characteristic features:
(a) a broad peak centered at $\sim$22 K above the SP transition,
(b) downward kink at $T_{SP}$ as temperature is
lowered, (c) unusual peak centered at $\sim$5.5 K below $T_{SP}$,
(d) strong suppression of the 5.5-K peak with high magnetic fields.
These observations are consistently understood by considering
both phonon and spinon heat transports, which are analyzed 
in combination with specific heat data. 
In particular, the unusual 5.5-K peak is a consequence of
phonons scattered both by defects and by spin excitations.

AK wishes to thank CRIEPI for the opportunity to spend time in their
laboratory in Komae where much of the work was done. 
SGD thanks the support of the Alexander von Humboldt Foundation. 
Work at Stanford University was
supported by AFOSR. 
Samples were prepared at the Center for Materials
Research, Stanford University.

%

\figure{FIG. 1.  Thermal conductivity and specific heat of CuGeO$_3$ single 
crystal in $H$=0.
Inset: Low temperatures diffusivity extracted by dividing 
$\kappa$ by the fitted lattice specific heat.
\label{fig1}}

\figure{FIG. 2.  A set of $\kappa(T)$ in magnetic fields up to 16 T.
Inset: $\kappa$ as a function of magnetic field at 4.2 K. 
Note the hysteresis at the SP to I transition at $\sim$12.5 T.
\label{fig2}}


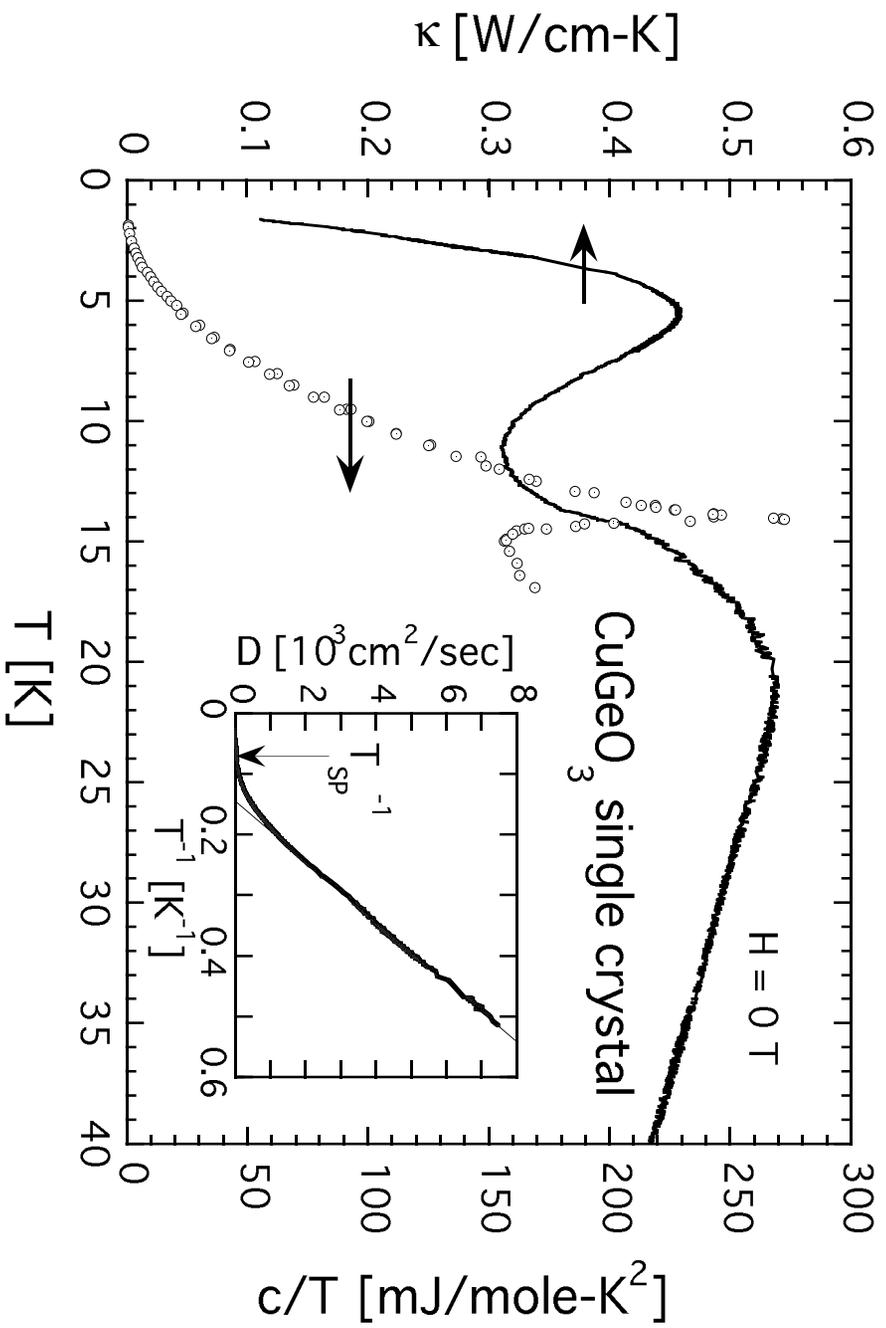
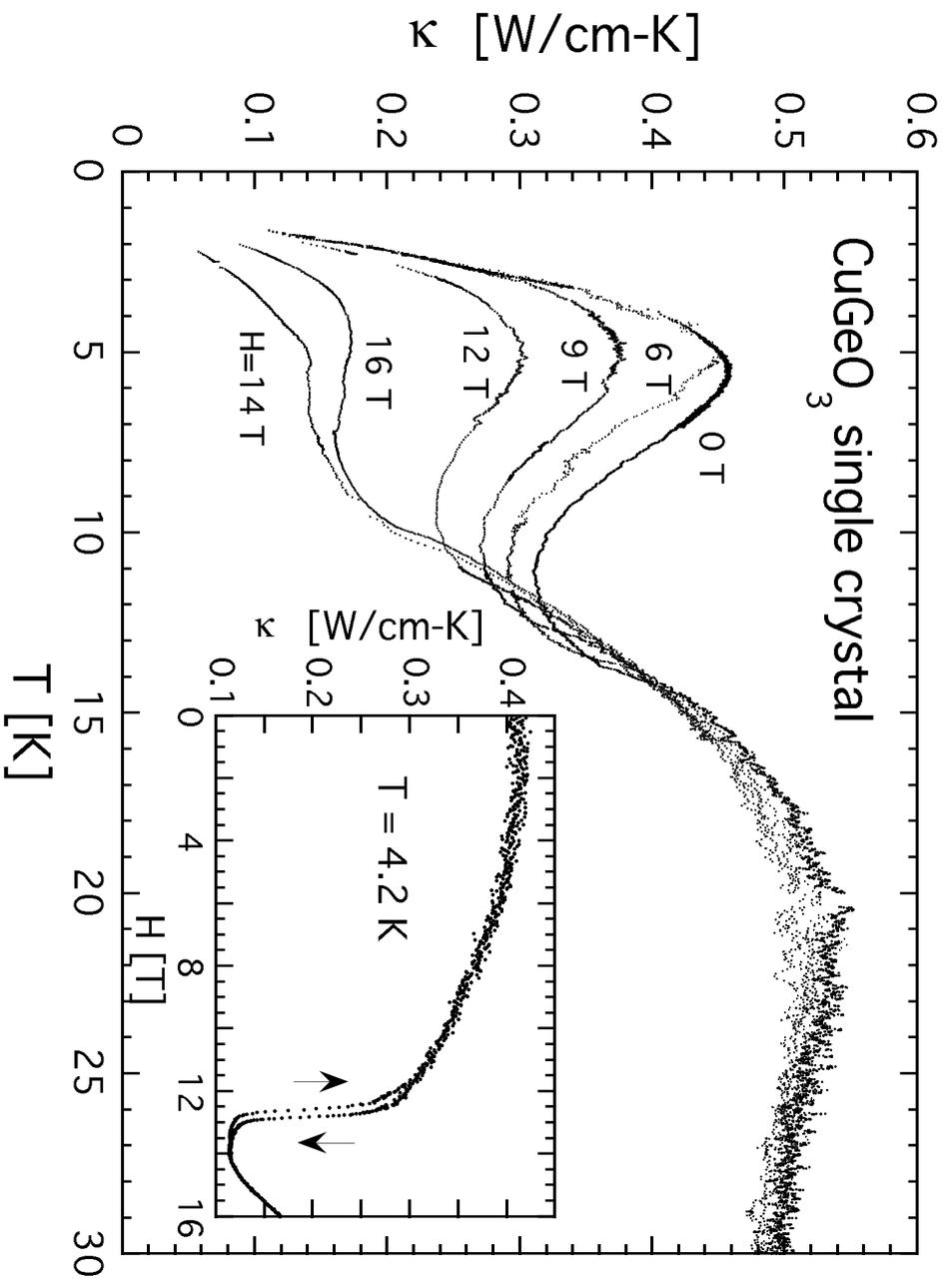
\end{document}